# HasTEE: Programming Trusted Execution Environments with Haskell

Abhiroop Sarkar
Chalmers University
Gothenburg, Sweden
sarkara@chalmers.se

Robert Krook
Chalmers University
Gothenburg, Sweden
krookr@chalmers.se

Alejandro Russo
Chalmers University
DPella AB
Gothenburg, Sweden
russo@chalmers.se

Koen Claessen
Chalmers University
Gothenburg, Sweden
koen@chalmers.se

## Abstract

Trusted Execution Environments (TEEs) are hardware enforced memory isolation units, emerging as a pivotal security solution for security-critical applications. TEEs, like Intel SGX and ARM TrustZone, allow the isolation of confidential code and data within an untrusted host environment, such as the cloud and IoT. Despite strong security guarantees, TEE adoption has been hindered by an awkward programming model. This model requires manual application partitioning and the use of error-prone, memory-unsafe, and potentially information-leaking low-level C/C++ libraries.

We address the above with *HasTEE*, a domain-specific language (DSL) embedded in Haskell for programming TEE applications. HasTEE includes a port of the GHC runtime for the Intel-SGX TEE. HasTEE uses Haskell's type system to automatically partition an application and to enforce *Information Flow Control* on confidential data. The DSL, being embedded in Haskell, allows for the usage of higher-order functions, monads, and a restricted set of I/O operations to write any standard Haskell application. Contrary to previous work, HasTEE is lightweight, simple, and is provided as a *simple security library*; thus avoiding any GHC modifications. We show the applicability of HasTEE by implementing case studies on federated learning, an encrypted password wallet, and a differentially-private data clean room.

## 1 Introduction

Trusted Execution Environments (TEEs) are an emerging design of hardware-enforced memory isolation units that aid in the construction of security-sensitive applications [Mulligan et al. 2021; Schneider et al. 2022]. TEEs have been used to enforce a strong notion of *trust* in areas such as confidential (cloud-)computing [Baumann et al. 2015; Zegzhda et al. 2017], IoT [Lesjak et al. 2015] and Blockchain [Bao et al. 2020]. Intel and ARM each have their own TEE implementations known as Intel SGX [Intel 2015] and ARM TrustZone [ARM 2004], respectively. Principally, TEEs provide a *disjoint* region of code and data memory that allows for the physical isolation of a program's execution and state from the underlying operating system, hypervisor, and I/O peripherals. For terminology, we shall use the term *enclave* (adopted from Intel) to refer to the isolated region of code and data and its trusted computing base (TCB).

TEEs, despite their strong security guarantees, have seen limited adoption in software development owing to several challenges. Firstly, TEEs often present *an awkward and low-level programming model* [Decentriq 2022]. For instance, Intel provides a C/C++ interface to program SGX that requires *partitioning* the program's state into trusted and untrusted components and dividing the entire logic into two separate software projects (Section 2)—a complex and error-prone process that could lead to data leakage. From a security perspective, the use of C/C++ APIs can open further opportunities to exploit well-known memory-unsafe vulnerabilities such as return-oriented programming (ROP) [Shacham 2007] in applications running inside TEEs [Muñoz et al. 2023]. Secondly, *current TEE programming models are insufficient to enforce security policies*. Applications should be written in a way such that they do not accidentally reveal confidential information. Furthermore, inputs and outputs to an enclave must be correctly encrypted, signed, decrypted, and verified to protect against malicious hosts. Thirdly, *little support is given to migrate legacy applications inside enclaves*. Applications inside enclaves often rely on their own Operating System (OS) since they cannot trust the one in the host machine. Library OS-based approaches exist to provide this functionality. However, for legacy applications written in high-level languages relying on non-trivial runtimes, the porting of the runtime becomes a challenging task.

Efforts have been made to address these challenges. The work by Ghosn et al. [2019] introduces GoTEE, a modification of the Go programming language with support for *secure routines* that are executed inside enclaves. In GoTEE, the authors heavily modify the Go compiler and extend the language to support new TEE-specific abstractions that helps to automatically partition an application. GoTEE does not provide any control over how sensitive information moves within the application, which could enable accidental data leaks. In a similar spirit, Oak et al. [2021] introduce $J_E$, a subset of Java with support for enclaves. $J_E$ focuses on providing information-flow control (IFC) to ensure that the code does not leak sensitive data by accident or by coercion of a malicious host. $J_E$ uses a sophisticated compilation pipeline to first partition the application and then uses another compiler to check that sensitive information is not leaked. Virtualization-based solutions, such as AMD SEV [AMD 2018], attempt to alleviate the effort required to port legacy applications. However, the trade-off is that the TCB becomes larger and the granularity to identify sensitive data becomes much coarser.

Our contribution through this paper is *HasTEE*, a domain-specific language (DSL) embedded in Haskell for programming TEE applications. HasTEE integrates TEE-specific abstraction and semantics while hiding low-level hardware intricacies making it hardware neutral! Additionally, HasTEE offers IFC to prevent accidental leakage of sensitive data. Owing to its embedding in Haskell, developers can use familiar abstractions such as high-order functions, monads, and a limited set of I/O operations to write applications in a conventional manner. This design choice enables seamless integration with all of the existing Haskell features. Compared to the previous work, HasTEE is lightweight, simple, and is provided as a *simple security library*; thus avoiding any GHC [Jones et al. 1993] compiler modifications!

## HasTEE by Example

Listing 1 presents a sample password checker application written using HasTEE.

```
1 pwdChkr :: Enclave String -> String -> Enclave Bool
2 pwdChkr pwd guess = fmap (== guess) pwd
3
4 passwordChecker :: App Done
5 passwordChecker = do
6   passwd <- inEnclaveConstant "secret"
7   efunc  <- inEnclave $ pwdChkr passwd
8   runClient $ do -- Client code
9     liftIO $ putStrLn "Enter your password"
10    userInput <- liftIO getLine
11    res       <- gateway (efunc <@> userInput)
12    liftIO $ putStrLn ("Login returned " ++ show res)
```

**Listing 1.** A password checker written in HasTEE

The distinction between the trusted and untrusted parts of the application is done via the type system that encodes the former as the `Enclave` type (line 1) and the latter as the `Client` type (type inferred in line 8).

The function `pwdChkr` takes a sensitive string located in the enclave (`Enclave String`), a public string from the client host (`String`) and produces a sensitive Boolean in the enclave (`Enclave Bool`). Line 6 holds the secret string that we want to protect (`inEnclaveConstant`). Line 7 uses the `inEnclave` call to obtain a reference to the function `pwdChkr` located in the enclave. The function `gateway` (line 11) is responsible for transmitting the collected arguments to the enclave function, and bringing the result back to the client. The `gateway` function acts as an interface between the enclave and non-enclave environment. The untrusted *host client* is in charge of driving the application, while the *enclave* is assigned the role of a computational and/or storage resource that services the client's requests. HasTEE connects an application (`passwordChecker`) to Haskell's `main` method using the `runApp :: App a -> IO a` function that executes the application. From an IFC perspective, lines 6 and 7 correspond to labelling, i.e., establishing, which inputs are sensitive for the program—an activity that is part of the TCB. In general, HasTEE code starts by labelling the sensitive input with the `inEnclave` primitive. Subsequently, the client code is compelled to manipulate secrets in a *secure* manner. In this setting, secure means that no sensitive information in the enclave gets leaked except that it has been obtained via the primitive `gateway`. The HasTEE API is explained in Section 4.2, and the semantics are discussed in Section 4.3.

### Contributions

*A type-safe, secure, high-level programming model.* The HasTEE library enables developers to program a TEE environment, such as Intel SGX, using Haskell - a type-safe, memory-managed language whose expressive type system can be leveraged to enforce various security constraints. Additionally, HasTEE allows programming in a familiar client-server style programming model (Section 4.2 and 5.2), an improvement over the low-level Intel SGX APIs.

**Automatic Partitioning.** A key part of programming TEEs, partitioning the trusted and untrusted parts of the program is done automatically using the type system (details in Section 3 and 4.3). Crucially, our approach does not require any modification of the GHC compiler and can be adapted to other programming languages, as long as their runtime can run on the desired TEE infrastructure.

**Information Flow Control.** Drawing inspiration from restricted IO monad families in Haskell, we designed an Enclave monad that prevents accidental leaks of secret data by TEE programmers (Section 5.3). Hence, our Enclave monad enables writing applications with a relatively low level of trust placed on the enclave programmer.

**Portability of Haskell's runtime.** We modify the GHC runtime, without modifying the compiler, to run on SGX enclaves. This enables us to host the complete Haskell language, including extensions, supported by GHC 8.8 (Section 5.1).

**Demonstration of expressiveness.** We illustrate the practicality of the HasTEE through three case studies across different domains: (1) a Federated Learning example (Section 6.1), (2) an encrypted password wallet (Section 6.2) and (3) a *differentially-private* data clean room (Section 6.3). The examples also demonstrate the simplicity of TEE development enabled by HasTEE.

## 2 Background

### Intel Software Guard Extensions (SGX)

Intel Software Guard Extensions (SGX) [Intel 2015] is a set of security-related instructions supported since Intel's sixth-generation Skylake processor, which can enhance the security of applications by providing a *secure enclave* for processing sensitive data. The enclave is a disjoint portion of memory separate from the DRAM, where sensitive data and code reside, beyond the influence of an untrusted operating system and other low-level software.

Intel offers an SGX SDK for programming enclaves. The SDK requires dividing the application into trusted and untrusted parts, where sensitive data resides in the trusted project. It provides specialized function calls called *ecall* for enclave access and an *ocall* API for communication with the untrusted client. The boundary between the client and enclave is defined using an *Enclave Description Language (EDL)*. The SDK utilizes a tool called *edger8r* to parse EDL files and generate two *bridge* files. These files ensure secure data transfer between projects through *copying* instead of sharing via pointers, preventing potential manipulation of the enclave's state. Fig 1 shows the SDK's programming model.

Application developers working with enclaves aim to minimize the Trusted Computing Base (TCB) by keeping the operating system and system software outside the enclave. The SGX SDK offers a restricted C standard library implementation (`tlibc`) for essential system software. Programming

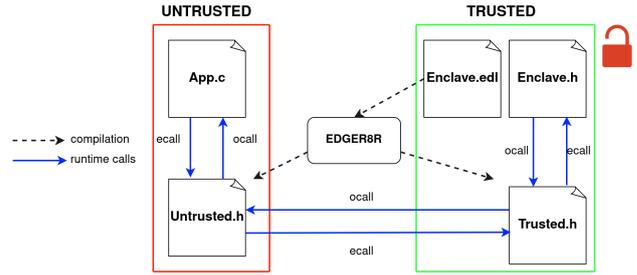

**Figure 1.** Intel SGX SDK Programming Model

SGX enclaves involves understanding the complex control flow between trusted and untrusted components. Enforcing SGX's programming model on typical software projects can be challenging, and the limited `tlibc` library restricts running applications beyond those written in vanilla C/C++.

## 3 Key Idea: A Typed DSL for Enclaves

### The Programming Model and Partitioning

HasTEE supports the automatic partitioning of programs by utilizing a combination of the type system to identify the enclave and a conditional compilation tactic to provide different semantics to each component. The compilation tactic was first used in Haste.App [Ekblad and Claessen 2014], to partition a single program into a `Client` and `Server` type. Fig 2 shows the partitioning procedure at a high level.

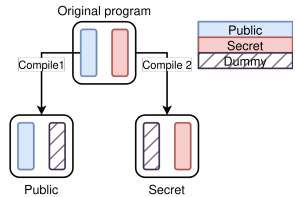

**Figure 2.** The HasTEE partitioning scheme

Importantly, this approach does not require any compiler extensions or elaborate dependency analysis passes to distinguish between the underlying types. The codebase involved in other complex partitioning approaches [Ghosn et al. 2019; Oak et al. 2021] becomes part of the Trusted Computing Base (TCB), creating a larger TCB. In contrast, our approach does not add any partitioning code to the TCB. Fig 3 shows the partitioned software stack in the HasTEE approach.

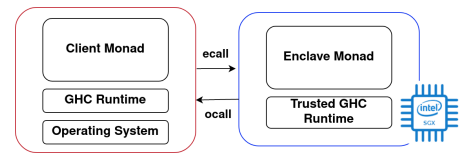

**Figure 3.** The untrusted (left) and trusted (right) software

Abhiroop Sarkar, Robert Krook, Alejandro Russo, and Koen Claessen

Post-partitioning, the client-server-style programming model is used for programming the enclave. In this model, the client takes on the primary role of driving the program and utilizes the enclave as a computational and/or storage resource. The source program, written in Haskell, benefits from type safety, while HasTEE internally handles the message transfer between the client and enclave memory at runtime.

**Information Flow Control on Enclaves**

Being a Haskell library enables HasTEE to tap into the library-based Information Flow Control techniques in Haskell [Buiras et al. 2015; Russo 2015; Russo et al. 2008]. The IFC literature distinguishes between sensitive and non-sensitive computations via monads indexed with security levels [Russo et al. 2008], e.g., Sec H and Sec L, where security levels H and L are assigned to sensitive and public information, respectively. Public information can flow into sensitive entities but not in the other way around. We have a similar security-level hierarchy between the Enclave and Client monads, respectively. Accordingly, we design the Enclave monad such that it restricts the possible variants of I/O operations. Internally, the Enclave monad constrains the scope of side-effecting operations to protect the confidentiality of data within the enclave (details in Section 5.3). Furthermore, HasTEE demands to explicitly mark where information is being sent back to the client (gateway), thus clearly indicating where to audit and control information leakages. Due to the security-critical nature of the Enclave monad, we include a trust operator, which is similar to the endorse function found in IFC literature.

**Trusted GHC Runtime**

One of the key challenges in allowing Haskell programs to run on TEE platforms is to provide support for the GHC Haskell Runtime [Marlow et al. 2009] itself. A Haskell program relies on the runtime for essential tasks such as memory allocation, concurrency, I/O management, etc. The GHC runtime heavily depends on well-known C standard libraries, such as glibc on Linux [GNUDevs 1991] and msvcrt on Windows [Microsoft 1994]. In contrast, the Intel SGX SDK provides a much more restricted libc known as tlibc.

This results in the fact that several libc calls used by the GHC runtime such as mmap, madvise, epoll, select and 100+ other functions become unavailable. Even the core threading library used by the GHC runtime, pthread, has a much more restricted API on the SGX SDK. To solve this conundrum, we have patched portions of the GHC runtime and used functionalities from a library OS, Gramine [C. Tsai, Porter, et al. 2017], to enable the execution of GHC-compiled programs on the enclave.

**TEE Independence**

Finally, HasTEE provides an abstraction over low-level system APIs offered by TEEs. As a result, the principles applied in programming Intel SGX should translate to the programming of other popular TEEs, such as the ARM TrustZone.

## 4 Design of HasTEE

### 4.1 Threat Model

We begin by discussing the threat model of the HasTEE DSL. HasTEE has the very same threat model as that of Intel SGX. In this model, only the software running inside the enclave memory is trusted. All other application and system software, such as the operating system, hypervisors, driver firmware, etc., are considered compromised by an attacker. A very similar threat model is shared by a number of other work based on Intel SGX [Arnautov et al. 2016; Baumann et al. 2015; Ghosn et al. 2019; Lind et al. 2017].

In this work, we enhance the application-level security firstly by using a memory-safe language, Haskell, and secondly use the Enclave monad to introduce information flow control. Our implementation strategy of loading the GHC runtime on the enclave allows us to handle Iago attacks [Checkoway and Shacham 2013] (see Section 5.1). We trust the underlying implementation of the SGX hardware and software stack (such as tlibc) as provided by Intel. Known limitations of Intel SGX such as denial-of-service attacks and side-channel attacks [Schaik et al. 2022] are beyond the scope of this paper.

An ideally secure development process should include auditing the code running on the enclave either through static analyses or manual code reviews or both. The conciseness of Haskell codebases should generally facilitate the auditing process. However, the mechanisms for fail-proof audits are beyond the scope of this paper as well.

### 4.2 HasTEE API

We show the core API of HasTEE in Fig 4. The functions presented operate over three principal Haskell data types: (1) Enclave, (2) Client, and (3) App. All three types are instances of the Monad typeclass, which allows for the use of do notation when programming with them. One of the key differences in functionality provided by the Client and Enclave monads is that Client allows for arbitrary I/O, whereas Enclave only provides restricted I/O. More on the latter in Section 5.3. The App monad sets up the infrastructure for communication between the Client and Enclave monad. We show a simple secure counter written using most of the API in Listing 2.

Listing 2 internally gets partitioned into the trusted and untrusted components via conditional compilation. In line 3, liftNewRef is used to create a secure reference initialised to the value 0. Followed by that, the computation to increment this value inside the enclave is given in lines 4 - 7. Applying inEnclave on the enclave computation (line 4) yields the type App (Secure (Enclave Int)). The Secure type is HasTEE's internal representation of a closure. Line

```haskell
-- mutable references
liftNewRef :: a → App (Enclave (Ref a))
readRef    :: Ref a → Enclave a
writeRef   :: Ref a → a → Enclave ()

-- get reference to function inside enclave
inEnclave :: Securable a ⇒ a → App (Secure a)

-- runs the Client monad
runClient :: Client () → App Done

-- used for function application on the enclave
gateway :: Binary a ⇒ Secure (Enclave a) → Client a
(<@>) :: Binary a ⇒ Secure (a → b) → a → Secure b

-- call this from 'main' to run the App monad
runApp :: App a → IO a
```

**Figure 4.** The core HasTEE API

```haskell
1  app :: App Done
2  app = do
3    enclaveRef <- liftNewRef 0 :: App (Enclave (Ref Int))
4    count <- inEnclave $ do
5      r <- enclaveRef
6      v <- readRef r
7      writeRef r (v + 1) >> return v :: Enclave Int
8    runClient $ gateway count >>=
9                  \v -> liftIO $ print $ "Counter's #" ++ show v
10
11 main = runApp app
```

**Listing 2.** A secure counter written in HasTEE (types annotated for clarity)

8 uses the critical `gateway` function to actually execute the enclave computation within the enclave memory and get the result back in the client memory. This resulting value, `v`, is displayed to the user.

The only function from Fig. 4 not used in Listing 2 is the `<@>` operator, used to collect arguments that are sent to the enclave. For example, an enclave function, `f`, that accepts two arguments, `arg1` and `arg2`, would be executed as `gateway (f <@> arg1 <@> arg2)`. Parameters to secure functions are copied to the enclave before the function is invoked, and results are copied from the enclave to the client before the client resumes execution. To do this copying, `gateway` and `<@>` has a `Binary` constraint on the types involved. This specifies that the values of the types involved have to be serialisable. Listing 1 in Section 1 shows a concrete usage of the operator. We have larger case studies in Section 6.

### 4.3 Operational Semantics of HasTEE

We provide big-step operational semantics of the HasTEE DSL. Note, we illustrate the semantics using an interpreter written in Haskell that shows the transition of the client as well as the enclave memory as each operators gets interpreted. We show our *expression language* and the abstract machine values to which we evaluate below:

```haskell
type Name = String

data Exp = Lit Int | Var Name | Fun [Name] Exp
  | App Exp [Exp] | Let Name Exp Exp | Plus Exp Exp
  | InEnclave Exp | Gateway Exp | EnclaveApp Exp Exp --HasTEE

data Value = IntVal Int | Closure [Name] Exp Env
  | SecureClosure Name [Value] | ArgList [Value] | Dummy
  | Err ErrState -- Error conditions
```

The `Exp` language above is a slightly modified version of lambda calculus with the restriction of allowing only fully applied function application. This restriction is done to reflect the nature of the HasTEE API, which through the type system, only permits fully saturated function applications for functions residing in the enclave. The lambda calculus language is then extended with the core HasTEE operators.

In the `Value` type, the `Closure` constructor, owing to saturated function application, captures a list of variable names and the environment. Notable in the `Value` type is the `SecureClosure` constructor that represents a closure residing in the enclave memory. This constructor does not capture the body of the closure as the body could hold any hidden state that lies protected within the enclave memory. The `SecureClosure` value is used by the `Gateway` function to invoke functions residing in the enclave.

The `ArgList` constructor supports the `<@>` operator that collects enclave function arguments. Lastly, the `Dummy` value is used as a placeholder for operators lacking semantics depending on the client or the enclave memory. For instance, the `Gateway` function has no meaning inside the `Enclave` monad, it is only usable from the `Client` monad. The `Dummy` crucially enables the conditional compilation trick in HasTEE by acting as a placeholder for meaningless functions in the respective client and enclave memory.

Our evaluators will show transition relations operating on two distinct memories that maps variable names to values - the enclave memory and the client memory.

```haskell
type ClientEnv  = [(Name, Value)]
type EnclaveEnv = [(Name, Value)]
```

Accordingly, we define two evaluators - `evalEnclave` (Fig. 5) and `evalClient` (Fig. 6). The complete evaluator run in two passes. In the first pass, it runs a program and loads up the necessary elements in the enclave memory and then in the second pass, the loaded enclave memory is additionally passed to the client's evaluator.

Two helper functions, `genEncVar` and `evalList` are not shown for concision. They generate unique variable names and fold over a list of expressions respectively. Appendix A contains the complete, typechecked semantics as runnable Haskell code.

We use Listing 3 to illustrate how the enclave, as well as the client memory, evolves as a program gets evaluated. Our

```
1  evalEnclave :: (MonadState StateVar m)
2              ⇒ Exp → EnclaveEnv → m (Value, EnclaveEnv)
3  evalEnclave (Lit n) env    = pure (IntVal n, env)
4  evalEnclave (Var x) env    = pure (lookupVar x env, env)
5  evalEnclave (Fun f) env =
6    pure (Closure xs e env, env)
7  evalEnclave (Let name e2) env = do
8    (e1', env') ← evalEnclave e1 env
9    evalEnclave e2 (name,e1'):env')
10 evalEnclave (App f args) env     = do
11   (v1, env1)  ← evalEnclave f env
12   (vals, env2) ← evalList args env1 []
13   case v1 of
14     Closure xs body ev →
15       evalEnclave body ((zip xs vals) ++ ev)
16     _ → pure (Err ENotClosure, env2)
17 evalEnclave (Plus e1 e2) env = do
18   (v1, env1) ← evalEnclave e1 env
19   (v2, env2) ← evalEnclave e2 env1
20   case (v1, v2) of
21     (IntVal a1, IntVal a2) → pure (IntVal (a1 + a2), env2)
22     _                    → pure (Err ENotIntLit, env2)
23 evalEnclave (InEnclave e) env = do
24   (val, env') ← evalEnclave e env
25   varname     ← genEncVar
26   let env''   = (varname, val):env'
27   pure (Dummy, env'')
28 -- the following two are essentially no-ops
29 evalEnclave (Gateway e) env = evalEnclave e env
30 evalEnclave (EnclaveApp e1 e2) env = do
31   (_, env1) ← evalEnclave e1 env
32   (_, env2) ← evalEnclave e2 env1
33   pure (Dummy, env2)
```

**Figure 5.** Operational Semantics of the Enclave

```
1  evalClient :: (MonadState StateVar m)
2              ⇒ Exp → ClientEnv → m (Value, ClientEnv)
3
4  -- evalClient for Lit, Var, Fun, Let, App, Plus not
5  -- shown as they behave the same as evalEnclave above
6  evalClient (InEnclave e) env = do
7    (_, env') ← evalClient e env
8    varname   ← genEncVar
9    let env'' = (varname, Dummy):env'
10   pure (SecureClosure varname [], env'')
11 evalClient (Gateway e) env = do
12   (e', env') ← evalClient e env
13   case e' of
14     SecureClosure varname vals → do
15       enclaveEnv ← gets encState
16       let func = lookupVar varname enclaveEnv
17       case func of
18         Closure vars body encEnv → do
19           (res,enclaveEnv') ←
20             evalEnclave body ((zip vars vals) ++ encEnv)
21           pure (res, env1)
22         _ → pure (Err ENotClosure, env1)
23     _ → pure (Err ENotSecClos, env1)
24 evalClient (EnclaveApp e1 e2) env = do
25   (v1, env1) ← evalClient e1 env
26   (v2, env2) ← evalClient e2 env1
27   case v1 of
28     SecureClosure f args →
29       case v2 of
30         ArgList vals →
31           pure (SecureClosure f (args ++ vals), env2)
32         v → pure (SecureClosure f (args ++ [v]), env2)
33     v → pure (ArgList [v,v2], env2)
```

**Figure 6.** Operational Semantics of the Client

```
1 testProgram = let m = 3 in
2               let f = λ x -> x + m in
3               let y = inEnclave f in
4               gateway (y <@> 2)
```

**Listing 3.** A simple program for illustrating the operational semantics of HasTEE

semantic evaluator operates in two passes. In the first pass, the evalEnclave evaluator from Fig. 5 is run. Fig. 7a shows the state of the enclave environment after the evaluator has completed evaluating Listing 3. Notably, the variable y maps to a value with no semantic meaning, as the evaluator is already running in the secure memory.

In the second pass, the environment from Fig. 7a is additionally passed as a state variable to the evaluator evalClient from Fig. 6. Note the different value mapped to the variable y in Fig 7b. EnclaveApp is evaluated on lines 25-34 in Fig 6. It generates the value SecureClosure "$EncVar_0$" [Lit 2].

Notable is the evaluation of the gateway call on line 4 of Listing 3. The semantics for this evaluation are in lines 12-24 of Fig 6. The evaluator upon finding a reference $EncVar_0$ with no semantics in the client memory (Fig 7b) looks up

$$
\begin{array}{ll}
m \mapsto 3 & m \mapsto 3 \\
f \mapsto Closure\ ["x"]\ (x+m)\ [m \mapsto 3] & f \mapsto Closure\ ["x"]\ (x+m)\ [m \mapsto 3] \\
EncVar_0 \mapsto Closure\ ["x"]\ (x+m)\ [m \mapsto 3] & EncVar_0 \mapsto Dummy \\
y \mapsto Dummy & y \mapsto SecureClosure\ "EncVar_0"\ []
\end{array}
$$

**(a)** Enclave Environment  **(b)** Client Environment

**Figure 7.** (a) gets loaded during the first evaluator pass, and the Client Environment remains empty. In the second pass, (b) gets loaded while having access to the memory (a), as can be seen in Fig 6.

$EncVar_0$ in the enclave environment (Fig 7a) and finds a Closure with a body. Crucially, **it evaluates the Closure by invoking the evalEnclave function on line 21 of Fig. 6 using the enclave environment.** This part models how the SGX hardware switches to the enclave memory when executing the secure function f rather than the client memory. An important point is generating an identical fresh variable name, $EncVar_0$, that the client uses to identify and call the functions in the enclave memory.

## 4.4 Practical security analysis

In what follows, we perform a security analysis of Has-TEE. We start by making explicit that the only communication from the enclave back to the host client is primitive `gateway`. In this regard, we have the following claim capturing a (progress-insensitive [Askarov, Hunt, et al. 2008]) non-interference property. Intuitively, this property states that (side-effectful) programs do not leak information except via their termination behavior.

**Proposition 4.1** (Non-interference). *Given a HasTEE program* `p :: Enclave a -> App Done`, *where* `p` *does not use primitive* `gateway`, *and two enclave computations* `e1 :: Enclave a` *and* `e2 :: Enclave a`, *then* `p e1` *and* `p e2` *perform the same side-effects in the host client.*

This proposition states that in `p` the public effects on the host client cannot depend on the content of the argument of type `Enclave a`. The veracity of this proposition can be proven from the semantics of `gateway`, which is the only primitive calling `evalEnclave` from `evalClient` Fig. 6. If non-interference does not hold in the context of developing HasTEE, it could indicate the presence of vulnerabilities in the system. For example, it could suggest that data is being leaked into the host environment due to an error in the partitioning process of the HasTEE compiler. Alternatively, it might imply that certain side effects within the enclave are unintentionally revealing data back to the host, contrary to our expectations. Non-interference serves as an important initial security condition in the development of HasTEE as it helps identify and address numerous vulnerabilities that may arise during the process.

When it comes to reason about programs with the primitive `gateway`, we need to reason about IFC *declassification* primitives (or intended ways to release sensitive information) [Sabelfeld and Sands 2005] and how to avoid exploiting it to reveal more information than intended. Gollamudi and Chong [2016] utilizes *delimited release* as the security policy. This security policy extends information-flow control beyond non-interference. It allows for explicit points of controlled information release, called *escape hatches*, where sensitive information can be sent to public channels. This policy stipulates that information may only be released through escape hatches and no additional information is leaked. The function `gateway` is our escape hatch. If we apply delimited release to HasTEE, then host clients can always learn what the function `gateway e` returns, given that expression `e` evaluates to the same value in the initial states `st1`, `st2 :: Enclave a` given to a program `p`—a condition to avoid misusing escape hatches to reveal more information than intended. Our case studies (Section 6) satisfy delimited release.

Automatically enforcing delimited release or robust declassification [Myers, Sabelfeld, et al. 2004] imposes severe restrictions in either the information being declassified or how declassification primitives are used. Hence, we leave enforcing such security policies as future work. Instead, our DSL explicitly requires marking the points where information is sent back to the client (i.e., `gateway`), making it clear where to audit and control information leakages.

## 5 Implementation of HasTEE

### 5.1 Trusted GHC Runtime

One of the crucial challenges in implementing the HasTEE library is enabling Haskell programs to run within an Intel SGX enclave. All Haskell programs compiled via the Glasgow Haskell Compiler (GHC), rely on the GHC runtime [Marlow et al. 2009] for crucial operations such as memory allocation and management, concurrency, I/O management, etc. As such, it is essential to port the GHC runtime in order to run Haskell programs on the enclave.

The GHC runtime is a complex software that is heavily optimized for specific platforms, such as Linux and Windows, to maximize its performance. For instance, on Linux, the runtime relies on a wide variety of specialised low-level routines from a C standard library, such as `glibc` [GNUDevs 1991] or `musl` [Felker 2005], to provide essential facilities like memory allocation, concurrency, and more. The challenge lies in porting the runtime due to the limited and constrained implementation of the C standard library in the SGX SDK, called `tlibc` [Intel 2018]. Specifically, `tlibc` does not support some of the essential APIs required by the GHC runtime, including `mmap`, `madvise`, `munmap`, `select`, `poll`, a number of `pthread` APIs, operations related to timers, file reading, writing, and access control, and 100+ other functions.

Given the magnitude of engineering effort required to port the GHC runtime, we fall back on a library OS called Gramine [C. Tsai, Porter, et al. 2017]. Gramine internally intercepts all `libc` system calls within an application binary and maps them to a Platform Abstraction Layer (PAL) that utilizes a smaller ABI. In Gramine's case, this amounts to only 40 system calls that are executed through dynamic loading and runtime linking of a larger `libc` library, such as `glibc` or `musl`. Importantly, to protect the confidentiality and integrity of the enclave environment, Gramine uses a concept known as *shielded execution*, pioneered by the Haven system [Baumann et al. 2015], where a library is only loaded if its hash values are checked against a measurement taken at the time of initialisation. Shielded execution further protects applications against *Iago attacks* [Checkoway and Shacham 2013] in Gramine.

However, there are additional difficulties in loading the GHC runtime on the SGX enclave via Gramine. Owing to Gramine's diminished system ABI, it has a dummy or incomplete implementation for several important system calls that the runtime requires. For instance, the absence of the `select`, `pselect`, and `poll` functions, which are used in the GHC IO manager, required us to modify the GHC I/O manager to manually manage the polling behavior through experimental heuristics. Similarly, the critical `mmap` operation in GHC uses specific flags (`MAP_ANONYMOUS`) that require modification. In addition, other calls, such as `madvise`, `getrusage`, and timer-based system calls, also require patching. We hope to quantify these modifications' performance in the future.

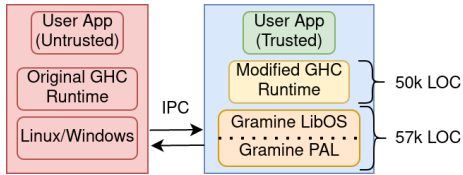

**Figure 8.** The high-level overview of communication between the untrusted and trusted parts of the app

After the GHC runtime is loaded onto an enclave, communication between the untrusted and trusted parts of the application effectively occurs between two disjoint address spaces. Communication between them can happen over any binary interface, emulating a remote procedure call. Our early prototype stage implementation uses an inter-process communication (IPC) call to copy the serialised data (Fig 8). A production implementation should communicate via the C ABI using Haskell's Foreign Function Interface (FFI), as this would be significantly faster than an IPC.

The Gramine approach requires 57,000 additional lines of code in the Trusted Computing Base (TCB) [C. Tsai, Porter, et al. 2017]. However, this is still an improvement over traditional operating systems, like Linux, with a TCB size of 27.8 million lines of code [Larabel 2020].

### 5.2 HasTEE Library

The API of the HasTEE library was already shown (Figure 4) and discussed in Section 4.2. The principal data types, `Enclave` and `Client`, have been implemented as wrappers around the IO monad, as shown below:

```
newtype Enclave a = Enclave (IO a) -- data constructor not
    exported

type Client = IO
```

A key distinction is that the `Enclave` data type does not instantiate the `MonadIO` typeclass, as a result of which arbitrary IO actions cannot be lifted inside the `Enclave` monad. This is to ensure that the enclave does not perform leaky IO operations such as writing to the terminal. These are effectful operations that may leak information, which may not be rolled back. However, the Enclave monad does instantiate a `RestrictedIO` typeclass that will be discussed in the following section. The conditional-compilation-based partitioning technique is achieved by having dummy implementations of certain data types in one of the modules, while the concrete implementation of those types is defined in the second module. We give an example of this using two different data types from the API.

```
-- Enclave.hs
data Secure a = SecureDummy

type Ref a = IORef a
```

```
-- Client.hs
data Secure a =
Secure CallID [ByteString]
type Ref a = RefDummy
```

A notable aspect of the API is the `Securable` typeclass, which constrains the `inEnclave` function and enables it to label functions with any number of arguments as residents of the enclave memory. The `Securable` typeclass accomplishes this using a well-known typeclass trick in Haskell, used to represent statically-typed variadic functions such as `printf` [Augustsson and Massey 2013]. In general, `Securable` characterises functions of the form $a_1 \rightarrow \dots \rightarrow a_n \rightarrow Enclave\ b$.

The operational semantics presented in Section 4.3 should provide an intuition for the core implementation techniques used in the library. The complete HasTEE project has been open-sourced[1]. More implementation details can be found in the Haste.App paper [Ekblad and Claessen 2014].

### 5.3 Information Flow Control for Enclaves

The HasTEE library, being written in Haskell, allows using language-based Information Flow Control (IFC) techniques available in Haskell [Russo et al. 2008]. IFC approaches in Haskell aim to protect the confidentiality of data by encapsulating computations within a `Sec` monad. Typically, the monad employs a lattice of *labels* [Denning 1976] to model various security levels and then enforces policies on how data can flow between the levels. For a two-label lattice, where confidential data is marked with H and public data with L, a security policy known as *non-interference* is to prevent information flow from the secret to public channels [Goguen and Meseguer 1982]. In other words, $L \sqsubseteq L, H \sqsubseteq H$, $L \sqsubseteq H$, but $H \not\sqsubseteq L$, where $\sqsubseteq$ indicates the *flows to* relation.

A similar scenario arises in HasTEE, where the `Enclave` monad can be compared to a security-critical `Sec` H monad that attempts to prevent information leakage to a public `Sec` L channel represented by the `Client` monad. Enforcing the non-interference policy in this scenario would imply that no data can flow out of the `Enclave` monad to the `Client`, which would make the enclave very restrictive for any real-world use cases. As such, the IFC literature relaxes the non-interference policy by the means of *declassification* [Sabelfeld and Sands 2005], to allow controlled data leak from H to L.

In the HasTEE API, the `gateway :: (Binary a) => Secure (Enclave a) -> (Client a)` function is an *escape hatch* [Hedin and Sabelfeld 2012] that allows the enclave to leak *any* data to the client. We prioritise the usability of the API and trust that the enclave programmer will make the gateway call when they are certain they want to intentionally leak information to a public channel. However, there is a hidden line of defence in the `gateway` function. If the programmer wishes to send any user-defined data type to the untrusted client, they need to provide an instance of the `Binary` typeclass. Writing this typeclass instance for some confidential data type, such as a private key, equips the confidential data with the capacity to leave the enclave boundary, which should be done in a highly controlled manner.

---

[1]

Besides the `gateway` function, the `Enclave` monad has occasional requirements to interact with general I/O facilities like file reading/writing or random number generation. For such operations, the `Enclave` monad would need a `MonadIO` instance in Haskell to perform any I/O operations. However, as discussed in the previous section, we do not provide the lenient `MonadIO` instance to the Enclave monad but instead, use a `RestrictedIO` typeclass to limit the types of I/O operations that an `Enclave` monad can do.

`RestrictedIO`, shown in Listing 4, is a collection of typeclasses that constrains the variants of I/O operations possible inside an `Enclave` monad. For instance, if a programmer, through the usage of a malicious library, mistakenly attempts to leak confidential data through a network call, the typeclass would not allow this.

```
1 type RestrictedIO m = (EntropyIO m, UnsafeFileIO m) -- other
      typeclasses not shown
2
3 class EntropyIO (m :: Type -> Type) where
4     type Entropy m:: Type
5     genEntropyPool :: m (Entropy m)
6
7 class UnsafeFileIO (m :: Type -> Type) where
8     untrustedReadFile :: FilePath -> m (Untrusted String)
```

**Listing 4.** The Restricted IO typeclass

This approach is invasive in that it restricts how a library (malicious or otherwise) that interacts with a HasTEE program conducts I/O operations. For instance, we had to modify the HsPaillier library [L.-T. Tsai and Sarkar 2016] that used the `genEntropy` function for random number generation. Initially, the library could use the Haskell IO monad freely, but to interact with a package written in HasTEE, it had to be modified to use the more restricted type class constraint (`EntropyIO`) for its *effectful* operations. This limits potential malicious behaviour within the library. Notably, our changes involve only five lines of code that instantiate the type class and generalize the type signature of effectful operations.

Another aspect of IFC captured in our system is the notion of *endorsement* [Hedin and Sabelfeld 2012], which is the dual of declassification. Endorsement is concerned with the integrity, i.e., trustworthiness, of information. In HasTEE, we utilize endorsement to ensure that the integrity of secrets is not compromised by data being introduced into the enclave.

HasTEE allows file reading operations inside the `Enclave` monad, which can potentially corrupt the enclave's data integrity. To control this, HasTEE provides two forms of file reading operation - (1) untrusted file read and (2) trusted encrypted file reads. For (1), data read from untrusted files require manual endorsement via the `trust :: Untrusted a -> a` operator (where `Untrusted a` is a wrapper over the data read). This provides an additional check before untrusted data interacts with the trusted domain.

For point (2), HasTEE relies on an Intel SGX feature known as *sealing*. Every Intel SGX chip is embedded with a unique 128 bit key known as the Root Seal Key (RSK). The SGX enclave can use this RSK to encrypt trusted data that it wishes to persist on untrusted media. This process is known as sealing; HasTEE provides a simple interface to seal as well as unseal the trusted data being persisted, as shown below:

```
1 data SecurePath = SecurePath String
2
3 securefile :: FilePath -> SecurePath
4 securefile fp = "/secure_location/" <> fp -- path hidden
5
6 readSecure  :: SecurePath -> Enclave String
7 writeSecure :: SecurePath -> String -> Enclave ()
```

In the above, the `writeSecure` operation corresponds to *ciphertext declassification* [Askarov, Hedin, et al. 2008], while `readSecure` to an operation that applies automatic endorsement if the file can be decrypted successfully by the enclave RSK. If an attacker were to locate the secure location, the worst possible outcome would be the deletion of the file. However, the contents of the file cannot be read or modified outside the enclave, so the attacker would not be able to access the sensitive information stored within.

## 6 Case Studies

### 6.1 Federated Learning

Federated Learning is an emerging *privacy-preserving* machine learning [Al-Rubaie and Chang 2019] approach that allows multiple parties to train a model without sharing the raw training data. A typical federated learning setup involves multiple decentralized edge devices holding local datasets, training a model locally and then aggregating the trained model on a cloud server. Fig. 9 shows the desired setup.

The setup in Fig. 9 above is facilitated by a combination of TEEs and homomorphic encryption. Homomorphic Encryption (HE) [Gentry 2009] is a form of encryption that enables direct computation on encrypted data, revealing the computation result only to the decryption key owner. We emulate the very same setup for our case study where we have two mutually distrusting parties -

• **Confidential data owner.** This party wants to protect its confidential data. A real-life example would be a hospital containing confidential patient data.

• **ML model owner.** This party wants to protect their intellectual property (the ML model) from the data owners as well as the cloud provider. They encrypt their model when sending it to the data owners and allows them to use only homomorphic encryption for operating on the model.

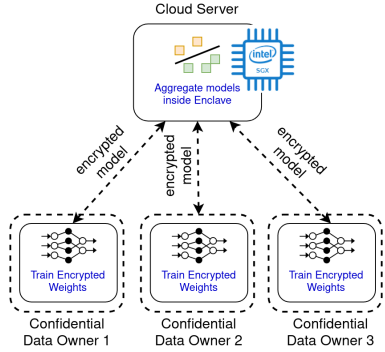

**Figure 9.** A Federated Learning setup where the data owners are protecting their data and the ML model owner is protecting their model. The training with encrypted weights can be done using homomorphic encryption.

```
1 data SrvSt =
2   SrvSt { publicKey :: PubKey, privateKey :: PrvKey
3         , updWts :: Vector Double, numClients :: Int
4         , wtsDict :: Map Epoch [Vector CipherText]) }
```

**Listing 5.** The Federated Learning server state

The above setup only requires the cloud server supporting Intel SGX technology so that even mobile devices can participate in training as a worker role. We can very conveniently model this entire setup as three clients and a server with an enclave in HasTEE. For illustration purposes, we will use GHC's threads to represent the three clients instead of three separate data owner machines.

Listing 5 models the server's state. Note that the weights are kept in plaintext form. The enclave state holds both its public and private keys. However, only the public key should be allowed to move to the client. We enforce this by not providing an instance of the `Binary` typeclass for the private key. If untrusted modules try to attack such enforcement by adding new instances to `Binary`, or even providing overlapping ones to override the behaviour of overloaded methods, then Safe Haskell [Terei et al. 2012] will indicate GHC to not compile the code. Haskell is unique in terms of having an extension like Safe Haskell. Safe Haskell enforces sandboxing for trusted code by banning extensions that introduce *loopholes* and compromise type-safety or module abstraction (often for the sake of performance). As discussed in Section 5.3, the lack of a `Binary` instance for the `privateKey` will prevent the enclave programmer from accidentally leaking the security-critical private key.

Listing 6 shows the API exposed to the client machine. Instead of the complex SGX_ECALL machinery, our API is expressed in idiomatic Haskell. Calling any function `f` from the record `api` with an argument `arg` in this API is expressed simply as `gateway ((f api) <@> arg)`.

```
1 type Accuracy = Double
2 type Loss     = Double
3 data API = API {
4   aggregateModel :: Secure (Epoch -> Vector CipherText ->
5     Enclave (Maybe (Vector CipherText))),
6   validateModel :: Secure (Enclave (Accuracy, Loss)),
7   getPublicKey :: Secure (Enclave PubKey),
8   reEncrypt :: Secure (CipherText -> Enclave CipherText)}
```

**Listing 6.** The Federated Learning client API

Listing 7 shows the main ML model training loop. A few functions have been elided for brevity, but the key portions of the client-server interaction in HasTEE should be visible. The `Config` type holds the state containing encrypted weights sent from the cloud server, the learning rate, the current epoch number and the public key. After each epoch it updates the weights to the new aggregated value (Line 12). The value `x'` is the data set that the data owners are protecting and `y` is the result of the learning algorithm. The `adjustModelWithLearningRate` function (body elided, line 6) takes the computed gradient (line 5) and tries to converge on the desired result.

On line 7 the server is communicated to aggregate models spread across different clients, with the `server` returning the encrypted updated weights `wt'`. We use a wrapper over `gateway`, called `retryOnEnclave` (body elided), to allow the server to move in lock step with all the clients. Then in line 8, the server is communicated again to collect the accuracy and loss in the ongoing epoch number, which gets displayed in line 9. Finally, the loop continues in line 10.

```
1 handleSingleEpoch :: API -> CurrentEpochNum -> MaxEpochNum
2   -> Matrix Double -> Vector Int -> Config -> Client
3   Config
4 handleSingleEpoch api n m x' y cfg'
5   | n == m    = return cfg'
6   | otherwise = do
7     grad   <- computeGradient api cfg' x' y
8     cfgNew <- adjustModelWithLearningRate api
9                   (cfg' { iterN = n }) grad
10    wt' <- retryOnEnclave $ (aggregateModel api) <@> n
11                   <@> (weights cfgNew)
12    (acc, loss) <- gateway (validateModel api)
13    printClient $ " Iteration no: " <> show n <>
14      " Accuracy: " <> show acc <> " Loss : " <> show loss
15    handleSingleEpoch api (n+1) m x' y
16                   (cfgNew { weights = wt' })
```

**Listing 7.** The key model training loop

Listing 7 above features a complex control flow with at least two interactions visible in the loop itself. Internally, `computeGradient` and `adjustModelWithLearning` both communicate with the enclave, calling the `reEncrypt` function to remove noise from the homomorphic encryption operation. HasTEE can represent a fairly complex, asynchronous control flow as simple Haskell function calls.

In terms of Information Flow Control, there are two important aspects in this case study. Firstly, the `RestrictedIO` typeclass constrains potentially malicious libraries from misbehaving. For example, consider the library HsPaillier [L.-T. Tsai and Sarkar 2016], which implements the Paillier Cryptosystem [Paillier 1999] for partial homomorphic encryption. All effectful operations from this library, such as `genKey :: Int -> IO (PubKey, PrvKey)`, need to be rewritten for them to be usable within the `Enclave` monad. The following snippet shows our typeclass instantiation and a sample type signature change needed inside the library.

```
1 instance (IO ~ m) => EntropyIO m where
2     type Entropy m = EntropyPool
3     genEntropyPool = createEntropyPool
4
5 -- genKey :: Int -> IO (PubKey, PrvKey) -- original type
6 genKey :: (Monad m, EntropyIO m) => Int -> m (PubKey, PrvKey)
```

The second aspect of IFC arises when the client machine queries the server for accuracy and loss by asking it to validate the model. On the server side, the enclave has to read a file with test data. This test data resides outside of the enclave and is potentially an attack vector. In order to not inadvertently trust such an exposed source, the enclave uses the `untrustedReadFile` function from the `RestrictedIO` typeclass (Listing 4). The file is read as an `Untrusted String` and requires explicit programmer *endorsement* via the `trust` operator for the compiler to typecheck the program.

Overall the case study constitutes only 500 lines of code. It naturally fits into the client-server programming model, and the usage of Haskell provides type safety and enables IFC-based security.

### 6.2 Encrypted Password Wallet

For this case study, we use HasTEE to implement a secure password wallet that stores authentication tokens in encrypted form on the disk. An authentication token can be retrieved from the wallet if the right master password is supplied. The definition of a password wallet used by the case study follows in Listing 8.

```
1 -- | A single entry of authentication tokens
2 data Item = Item { title :: String, username :: String,
        password :: Password } deriving (Show, Read)
3 -- | The secure wallet
4 data Wallet = Wallet { items :: [Item], size :: Int,
        masterPassword :: Password} deriving (Show, Read)
```

**Listing 8.** The definition of a password wallet as a regular Haskell data type.

The `Show` and `Read` instances are used to convert a wallet to and from a string. This allows us to write the wallet to disk, and by writing to a secure file path we ensure that the stored wallet is encrypted, as described in section 5.3. By omitting a `Binary` instance we ensure that the wallet is not inadvertently leaked to the client directly.

```
1 -- | Secure file path to the wallet
2 wallet :: SecureFilePath
3 wallet = secureFile "wallet.seal"
4
5 -- | Try to load the secure wallet into the enclave
6 loadWallet :: Enclave (Maybe Wallet)
7 loadWallet = do b <- doesSecureFileExist wallet
8                 if b then do contents <- readSecure wallet
9                         return $ readMaybe contents
10                     else return Nothing
11
12 -- | Store the wallet on disk in encrypted form
13 saveWallet :: Wallet -> Enclave ReturnCode
14 saveWallet w = writeSecure wallet (show w) >> return Success
```

**Listing 9.** The code that storing and loading the encrypted wallet. Programmer do not need to manage encryption keys.

in Listing 9 implements the functions that store and load the wallet. We emphasize that the code does not need to explicitly reason about encryption and decryption, except for defining the secure file path.

Our password wallet has the following features - (1) adding an authentication token, (2) retrieving a password, (3) deleting a token and (4) changing the master password. It is designed as a command-line utility where the commands are handled by an untrusted client and the passwords are protected by the enclave. The complete implementation is roughly 200 lines of Haskell code.

The hardware-enforced security provided by our secure wallet makes it a natural fit for designing wallets that are protected by biometrics. A similar approach is used on modern iPhones, where passwords are stored in a secure enclave [Apple 2021] to ensure confidentiality, and the user's biometric data is used as the master password. In our case, the usage of a high-level language like Haskell enables expressing this relatively complex application concisely.

### 6.3 Data Clean Room with Differential Privacy

A *Data Clean Room* (DCR) [AWS 2022] is a technology that provides aggregated and anonymised user information to protect user privacy while providing advertisers and analytic firms with non-personally identifiable information to target a specific demographic with advertising campaigns and analytics-based services.

DCRs compute and release aggregated results based on the user data. To prevent attackers from compromising individual user information from aggregate data (via statistical techniques), DCRs employ *differential privacy* [Dwork 2006]. Differential privacy adds calibrated noise to the aggregate data making it computationally hard for attackers to compromise individual data. The noise calibration can be adjusted for increased privacy (more noise) or increased accuracy (less noise).

Our third case study implements a *differentially-private* DCR within an SGX enclave using HasTEE. The DCR consists of record, `User`, containing fields such as `name`, `occupation`, `salary`, `gender`, `age`, etc. The `User` record is encrypted before being provisioned to the DCR, after which we use the *Laplace Mechanism* [Dwork and Roth 2014] when performing counting queries to add noise to the result. The mechanism introduces noise by sampling a Laplace distribution. The code implementing the Laplace mechanism can be found in Appendix B.

The DCR does not provide a `Binary` instance for the `User` record to ensure that it is not transferred to the enclave via plain serialisation. Instead, we expose functions that encrypt and decrypt users.

The Laplace Mechanism for adding noise requires a source of randomness. Here, we use Haskell's `System.Random` package, which internally reads from `/dev/urandom`. For production environments, a more cryptographically secure source of randomness is required. We extend the `RestrictedIO` (Section 5.3) interface to allow this operation as long as the programmer *endorses* the file read.

Consider a sample query to test how many individuals in a data set have a salary within a specific range.

```haskell
salaryWithin :: Integer -> Integer -> User -> Bool
salaryWithin l h u = l <= salary u && salary u <= h
```

The HasTEE code for the DCR executing this query is shown in Listing 10. Lines 3 to 8 specify the API of the data clean room. The DCR's API supports (1) initialisation, (2) fetching of the public key, (3) provisioning user data to the enclave, and (4) executing the salary query. Line 8 is used to generate some arbitrary users (for testing), after which the client code takes over. The client initializes the DCR and fetches its public key. After this, the users are encrypted and sent to the DCR. On line 15 the salary query is executed in the DCR, and then the result is printed.

Generating arbitrary users to test the setup is done purely for illustration purposes. In a more faithful implementation, the client would relay the public key to data owners that would then send already encrypted user records to the client, which provisions them to the DCR. Owing to HasTEE's client-server programming model and the use of a high-level language like Haskell, the implementation becomes very compact with roughly 200 LOC.

# 7 Evaluations

## 7.1 Discussion

In contrast to development on the Intel C/C++ SGX SDK, HasTEE's high-level programming model entirely abstracts away the complexity of dealing with the low-level `edl` files in the SGX SDK. The remote procedure calls that happen between the untrusted client and trusted enclave are typechecked in Haskell, unlike the SGX SDK. The benefits of high-level of abstraction can also be seen in the password wallet example,

```haskell
app :: App Done
app = do
  ref     <- liftNewRef undefined
  initSt  <- inEnclave $ initEnclave ref 0.1
  pkey    <- inEnclave $ getPublicKey ref
  prov'   <- inEnclave $ provisionUserEnclave ref
  lm      <- inEnclave $ laplaceMechanism ref $
                           salaryWithin 10000 50000
  dataset <- liftIO $ sequence $ replicate 500
                           (generate arbitrary)
  runClient $ do
    gateway $ initSt       -- initialize enclave state
    key <- gateway pkey    -- enclaves public key
    mapM_ (\u -> do ct <- encryptUser u key
                    gateway $ prov' <@> ct) dataset
                           -- provision users
    result <- gateway lm   -- run the salary query
    liftIO $ putStrLn $ concat ["res: ", show result]
```

**Listing 10.** The client running the salaryWithin query over the data set in the data clean room.

where functions `readSecure` and `writeSecure` (Listing 9) relieves developers from the burden of key management. Furthermore, HasTEE warns a program against accidental data leaks and can enforce stronger compile-time guarantees than Intel C/C++ SGX SDK. For instance, in all three case studies, the lack of the `Binary` type-class constraint would, by construction, prevent accidental leakage of the secret data from the enclave. All three case studies restrict the I/O operations possible in the `Enclave` monad by the type-class `RestrictedIO`. Notably, in the federated learning example, we adapted the homomorphic encryption library to limit the effects possible in the `IO` monad.

## 7.2 Performance Evaluations

Our evaluations were conducted on an Azure Standard DC1s v2 (1 vcpu, 4 GiB memory) SGX machine. We use the password wallet case study as the canonical example to present performance evaluations across different parameters. We chose this example as it covers all the major aspects of the HasTEE API, such as protecting the confidentiality of data across the memory as well as the disk.

**Memory Overhead.** We show the memory consumption of our modified GHC runtime, sampled across 100 runs, where a sample was collected every second.

| Memory  | RSS        | Virtual Size | Disk Swap |
|---------|------------|--------------|-----------|
| At rest | 19,132 KB  | 287,920 KB   | 0 KB      |
| Peak    | 20,796 KB  | 290,032KB    | 0 KB      |

Although the memory usage of HasTEE will certainly vary across applications, these numbers provide a general estimate of the trusted GHC runtime's space usage. The Resident Set Size (RSS) indicates that the application fits within 20 MB at peak usage. RSS is an overestimate of memory usage as it includes the memory occupied by shared libraries as well. As a result, we can be certain that our application fits within the Enclave Page Cache limit (Section 2) of 93 MB.

**Latency.** We measure the latency and throughput for an instance of password retrieval, that includes - (i) an enclave crossing to call the trusted runtime, (ii) standard GHC execution time, (iii) encrypted file load, (iv) file decryption, (v) file read, and (v) a second enclave crossing to return the result.

Our measurements show that using the Linux `send`/`recv` call for enclave crossing results in a *60 milliseconds overall latency*. As our current socket-based communication is a proof-of-concept, it incurs a substantial overhead compared to native SGX enclave crossings. As a baseline, we measured the latency of an encrypted password retrieval in unmodified GHC (file encrypted with gpg [GNU 1999]). The baseline number comes out to be 0.6 milliseconds showing an overall 100x slowdown. Note that an average SGX ECALL operation incurs at least a 10x slowdown via the native SDK [Ghosn et al. 2019]. We believe switching to native ECALLs has the potential to improve our latencies.

**Throughput.** In terms of throughput, HasTEE is able to handle on average 11 requests for password retrieval per second. Again, this number has the potential for further improvemnt by switching to native SGX ECALLs.

We currently present coarse-grained measurements of the various metrics but envision future work, where more fine-grained parameters, such as the correlation between the GC pauses across the two runtimes can be presented. Section 7.3 provides a qualitative comparison of HasTEE against GoTEE and $J_E$.

### 7.3 Comparing HasTEE to GoTEE and $J_E$

Table 1 presents a comparison between HasTEE and its two closest counterparts - GoTEE [Ghosn et al. 2019] and $J_E$ [Oak et al. 2021]. While both GoTEE and $J_E$ had to modify the respective compilers, HasTEE required no modifications to the compiler. The specific runtime used by $J_E$ is not mentioned in the paper [Oak et al. 2021]; however, it suggests that no modification of the runtime was required, as it was run on a large virtualized host - SGX-LKL [Priebe et al. 2019]. In contrast, the runtimes for HasTEE and GoTEE required modification. GoTEE required significant modifications to the Golang runtime system to enable communication between the trusted and untrusted memory.

Both GoTEE and $J_E$ use sophisticated static analysis passes and program transformations to partition a program into its two components. In contrast, HasTEE's conditional compilation-based approach is much simpler, which is beneficial when it comes to security. Having less and simpler code makes it easier to verify for correctness. Notably, the purity of Haskell enables the user to inspect the type of a function and infer that it is naturally confined whenever a function is side-effect free. Inferring the confinement of a pure function is much more challenging in imperative languages like Java and Go.

## 8 Related Work

**Managed programming languages.** While there are imperative and object-oriented languages with TEE support (e.g., Go-based [Ghosn et al. 2019], and Java-based[Oak et al. 2021; C. Tsai, Son, et al. 2020], HasTEE is (to the best of our knowledge) the first functional language running on a TEE environment. The Rust-SGX [Wang et al. 2019] project provides foreign-function interface (FFI) bindings to the C/C++ Intel SGX SDK. Different from HasTEE, Rust-SGX does not aim to introduce any programming model or IFC to protect against leakage of sensitive data. Instead, Rust-SGX's main goal is application-level memory safety when programming with the low-level SGX SDK. HasTEE provides memory safety by the virtue of running Haskell, a memory-safe language, on the enclaves. TrustJS [Goltzsche et al. 2017] takes a similar FFI-based approach as Rust-SGX for programming enclaves with JavaScript. An important project in this space is the WebAssembly (WASM) initiative [Rossberg 2019]. There have been WASM projects, both academic, such as Twine [Ménétrey et al. 2021], as well as commercial, such as Enarx [Red Hat 2019], aimed at allowing WASM runtimes to operate within SGX enclaves. Our initial approach was to use the experimental Haskell WASM backend [Tweag.io 2022] to run Haskell on SGX enclaves. However, the aforementioned runtimes are not supported by GHC and lack several key features required for loading Haskell onto an enclave.

**Automatic partitioning.** HasTEE has a seamless program partitioning and familiar client-server-based programming model for enclaves. HasTEE's lightweight partitioning approach is inspired by the Haste.App library [Ekblad and Claessen 2014]—a library to write web applications in Haskell and deploy parts of it into JavaScript on the web browser. The most well-known automatic partitioning tool for C programs on an SGX enclave is Glamdring [Lind et al. 2017]. The general idea of partitioning a single program has been studied as multitier programming [Weisenburger et al. 2021]. Among the existing approaches to multitier programming, HasTEE provides a lightweight alternative that does not require any compiler modification or elaborate dataflow analysis to partition the program.

**Application development.** There have been attempts to virtualize entire platforms within the enclave memory to reduce the burden of dealing with the two-project programming model of Intel SGX. Haven [Baumann et al. 2015] virtualizes the entire Windows operating system as well as an entire SQL server application running on top of it. SCONE [Arnautov et al. 2016] virtualizes a Docker container instance within an SGX enclave. The libraryOS Gramine [C. Tsai, Porter, et al. 2017], which is used in this work, is an example of lightweight virtualization.

AMD's TEE system, AMD SEV [AMD 2018], is natively a virtualization-based approach. While it eases development, virtualization can result in drastically increasing the size of the TCB. We chose to apply a libraryOS approach for HasTEE in order to have a TCB of 57k lines of code (Gramine). As a future work, we can move away from Gramine and make the GHC runtime a standalone library inside the SGX enclave.

| Framework | HasTEE | GoTEE | $J_E$ |
|---|---|---|---|
| IFC support | Standard declassification | None | Robust declassification |
| Partitioning scheme | Type-based | Process-based | Annotation-based |
| Modified compiler | No | Yes | Yes |
| Modified runtime system | Yes | Yes | No |
| Trusted Components | GHC compiler, GHC runtime, Gramine | GoTEE compiler, GoTEE runtime | Java parser and partitioner, Jif compiler, JVM, SGX-LKL[Priebe et al. 2019] |
| Programming model | Client-server | Synchronous Message-Passing | Using the object-framework provided by Java |

**Table 1.** Comparison of HasTEE, GoTee, and $J_E$. We specify the core components involved in the Trusted Computing Base in all three frameworks.

**Information Flow Control.** HasTEE draws inspiration from the work on static IFC security libraries (e.g., [Buiras et al. 2015; Russo 2015; Russo et al. 2008]). Such approaches relies on the purity of Haskell to detect and stop malicious behaviour. HasTEE can support IFC in a dynamic manner [Stefan et al. 2011] by adapting the interpretation of the Enclave type to be a runtime monitor rather than just a wrapper to IO, where gateway performs security checks when sending/receiving information—an interesting direction for future work.

The work on IMP$_E$ [Visser and Smaragdakis 2016] studies IFC non-interference for passive and active attackers on the host client. Gollamudi, Chong, and Arden [2019] present a calculus for reasoning about IFC for applications distributed across several enclaves. $J_E$ [Oak et al. 2021] studies how compromised host clients can abuse gateway (declassification) primitives. Their security property and enforcement is based on the notion of robust declassification [Myers, Sabelfeld, et al. 2004; Waye et al. 2015]. Intuitively, this policy ensures that low-integrity data cannot influence the declassification of secret data. HasTEE enforces a simpler IFC policy for passive attackers—along the lines of Visser and Smaragdakis [2016]—and defer automatic analyses of the use of gateway for future work. Another interesting line of work is Moat [Sinha et al. 2015], which formally verifies enclave programs running on Intel SGX such that data confidentiality is respected. It uses IFC to enforce the policies and automated theorem proving to verify the policy enforcement mechanism.

## 9 Conclusion & Future Work

This paper presents HasTEE, a domain-specific language to write TEE programs while ensuring confidentiality of data by construction. Unlike previous work, HasTEE provides its partitioning of source code and IFC as a library! For HasTEE to work, we ported GHC's runtime to run within SGX enclaves by using the libraryOS Gramine. We demonstrate through three diverse case studies how HasTEE's IFC mechanism can help prevent accidental data leakage while producing concise code. We hope HasTEE opens future research avenues at the intersection of TEEs and functional languages.

There are several directions for future work. The IFC scheme we consider operates on two security levels - sensitive (Enclave) and public (Client) data. A natural extension is to enable multiple security levels [Myers and Liskov 2000; Stefan et al. 2011] to represent the concerns of different principals contributing data to enclaves. TEEs also provide a verifiable launch of the execution environment for the sensitive code and data, enabling a remote entity to ensure that it was set up correctly. *Remote attestation* [Knauth et al. 2018] allows an SGX enclave to prove its identity to a challenger using the private key embedded in the enclave. HasTEE does not capture attestation at the programming language level since it a property of the system components layout. Nevertheless, remote attestation can facilitate secure communication between multiple enclaves, e.g., a distributed-enclave setting; so it would be interesting to incorporate language-level support for remote attestation. Finally, GHC runtime is extensively optimized for performance. Obtaining a more compact and portable runtime, e.g., by using a restricted set of `libc` operations, could result in a considerably smaller TCB. A more portable runtime would facilitate HasTEE experiments on other TEE infrastructures such as ARM TrustZone and RISC-V PMP [RISC-V 2017].

## Acknowledgements

This work was funded by the Swedish Foundation for Strategic Research(SSF) under the project Octopi (Ref. RIT17-0023) and VR. Special thanks to Mary Sheeran, Marco Vassena, Bo Joel Svensson, and Claudio Agustin Mista for their initial reviews, and Facundo Dominguez and Michael Sperber for guiding the published draft.

# A Typechecked Operational Semantics of HasTEE in Haskell

```haskell
1  {-# LANGUAGE FlexibleContexts #-}
2  module HasTEEOrig where
3
4  import Control.Monad.State.Class
5  import Control.Monad.State.Strict
6
7  type Name = String
8
9  data Exp = Lit Int
10          | Var Name
11          | Fun [Name] Exp
12          | App Exp [Exp]
13          | Let Name Exp Exp
14          | Plus Exp Exp
15
16          -- HasTEE operators
17          | InEnclave  Exp
18          | Gateway  Exp
19          | EnclaveApp Exp Exp -- (<@>)
20          deriving (Show)
21
22  data Value = IntVal Int
23           | Closure [Name] Exp Env
24           -- HasTEE values
25           | SecureClosure Name [Value]
26           | ArgList [Value]
27           | Dummy
28
29           -- Error values
30           | Err ErrState
31           deriving (Show)
32
33  data ErrState = ENotClosure
34              | EVarNotFound
35              | ENotSecClos
36              | ENotIntLit
37
38  instance Show ErrState where
39    show ENotClosure  = "Closure not found"
40    show EVarNotFound = "Variable not in environment"
41    show ENotSecClos  = "Secure Closure not found"
42    show ENotIntLit   = "Not an integer literal"
43
44  type Env = [(Name, Value)]
45
46  type ClientEnv  = Env
47  type EnclaveEnv = Env
48
```

```haskell
49
50  type VarName = Int
51
52
53  data StateVar =
54    StateVar { varName  :: Int
55             , encState :: EnclaveEnv
56             }
57
58  initStateVar :: EnclaveEnv -> StateVar
59  initStateVar = StateVar 0
60
61
62  eval :: Exp -> Value
63  eval e =
64    let newEnclaveEnv = snd $
65                          evalState (evalEnclave e
66      initEnclaveEnv)
                            (initStateVar initEnclaveEnv)
67    in fst $ evalState (evalClient e initClientEnv) (
      initStateVar newEnclaveEnv)
68    where
69      initEnclaveEnv = []
70      initClientEnv = []
71
72  genEncVar :: (MonadState StateVar m) => m String
73  genEncVar = do
74    n <- gets varName
75    modify $ \s -> s { varName = 1 + n }
76    pure ("EncVar" <> show n)
77
78  evalList :: (MonadState StateVar m) => [Exp] -> Env -> [
      Value] -> m ([Value], Env)
79  evalList []      e vals = pure (reverse vals, e)
80  evalList (e1:es) env xs = do
81    (v, e) <- evalEnclave e1 env
82    evalList es e (v:xs)
83
84
85  evalEnclave :: (MonadState StateVar m)
86              => Exp -> EnclaveEnv -> m (Value, EnclaveEnv)
87  evalEnclave (Lit n) env = pure (IntVal n, env)
88  evalEnclave (Var x) env = pure (lookupVar x env, env)
89  evalEnclave (Fun xs e) env =
90    pure (Closure xs e env, env)
91  evalEnclave (Let name e1 e2) env = do
92    (e1', env') <- evalEnclave e1 env
93    evalEnclave e2 ((name,e1'):env')
94  evalEnclave (App f args) env = do
95    (v1, env1) <- evalEnclave f env
96    (vals, env2) <- evalList args env1 []
97    case v1 of
98      Closure xs body ev ->
99        evalEnclave body ((zip xs vals) ++ ev)
100     _ -> pure (Err ENotClosure, env2)
101 evalEnclave (Plus e1 e2) env = do
102    (v1, env1) <- evalEnclave e1 env
103    (v2, env2) <- evalEnclave e2 env1
```

```
104  case (v1, v2) of
105    (IntVal a1, IntVal a2) -> pure (IntVal (a1 + a2), env2)
106    _ -> pure (Err ENotIntLit, env2)
107
108 evalEnclave (InEnclave e) env = do
109   (val, env') <- evalEnclave e env
110   varname      <- genEncVar
111   let env'' = (varname, val):env'
112   pure (Dummy, env'')
113 -- the following two are the essentially no-ops
114 evalEnclave (Gateway e) env = evalEnclave e env
115 evalEnclave (EnclaveApp e1 e2) env = do
116   (_, env1) <- evalEnclave e1 env
117   (_, env2) <- evalEnclave e2 env1
118   pure (Dummy, env2)
119
120 evalList2 :: (MonadState StateVar m) => [Exp] -> Env -> [
       Value] -> m ([Value], Env)
121 evalList2 []       e vals = pure (reverse vals, e)
122 evalList2 (e1:es) env xs = do
123   (v, e) <- evalClient e1 env
124   evalList2 es e (v:xs)
125
126
127 evalClient :: (MonadState StateVar m)
128            => Exp -> ClientEnv -> m (Value, ClientEnv)
129 evalClient (Lit n) env = pure (IntVal n, env)
130 evalClient (Var x) env = pure (lookupVar x env, env)
131 evalClient (Fun xs e) env =
132   pure (Closure xs e env, env)
133 evalClient (Let name e1 e2) env = do
134   (e1', env) <- evalClient e1 env
135   evalClient e2 ((name,e1'):env')
136 evalClient (App f args) env = do
137   (v1, env1) <- evalClient f env
138   (v2, env2) <- evalList2 args env1 []
139   case v1 of
140     Closure xs body ev ->
141       evalClient body ((zip xs v2) ++ ev)
142     _ -> pure (Err ENotClosure, env2)
143 evalClient (Plus e1 e2) env = do
144   (v1, env1) <- evalClient e1 env
145   (v2, env2) <- evalClient e2 env1
146   case (v1, v2) of
147     (IntVal a1, IntVal a2) -> pure (IntVal (a1 + a2), env2)
148     _ -> pure (Err ENotIntLit, env2)
149
150
151 evalClient (InEnclave e) env = do
152   (_, env') <- evalClient e env
153   varname      <- genEncVar
154   let env'' = (varname, Dummy):env'
155   pure (SecureClosure varname [], env'')
156 evalClient (Gateway e) env = do
157   (e', env1) <- evalClient e env
158   case e' of
159     SecureClosure varname vals -> do
160       enclaveEnv <- gets encState
```

```
161     let func = lookupVar varname enclaveEnv
162     case func of
163       Closure vars body encEnv -> do
164         (res,enclaveEnv') <- evalEnclave body ((zip vars
         vals) ++ encEnv)
165         pure (res, env1)
166       _ -> pure (Err ENotClosure, env1)
167   _ -> pure (Err ENotSecClos, env1)
168 evalClient (EnclaveApp e1 e2) env = do
169   (v1, env1) <- evalClient e1 env
170   (v2, env2) <- evalClient e2 env1
171   case v1 of
172     SecureClosure f args ->
173       case v2 of
174         ArgList vals -> pure (SecureClosure f (args ++ vals)
         , env2)
175         v -> pure (SecureClosure f (args ++ [v]), env2)
176     v -> pure (ArgList [v,v2], env2)
177
178
179 -- gateway (f == SecureClosure f [])
180 -- gateway (f <@> arg == EA f arg == SecureClosure f [arg])
181 -- gateway (f <@> arg1 <@> arg2 == EA f (EA arg1 arg2) == SC
       f [arg1, arg2])
182
183
184
185
186 lookupVar :: String -> [(String, Value)] -> Value
187 lookupVar _ [] = Err EVarNotFound
188 lookupVar x ((y, v) : env) =
189   if x == y then v else lookupVar x env
```

**Listing 11.** Operational Semantics of HasTEE

# B  Data Clean Room Code

```
1 -- This function runs the query over the dataset and
2 -- returns the true result
3 countingQuery :: Enclave (Ref CleanRoomSt) -> (User -> Bool)
       -> Enclave Int
4 countingQuery refst q = do
5   st <- readRef =<< refst
6   return $ length $ filter id $ map q (users st)
7
8 -- Sample the laplace distribution
9 laplaceDistribution :: Enclave (Ref CleanRoomSt) -> Double
       -> Enclave Double
10 laplaceDistribution refst b = do
11   z <- int2Double <$> getRandom refst (0,1)
12   u <- ((/) 1000 . int2Double) <$> getRandom refst
       (1,1000)
13   return $ (2 * z - 1) * (b * log u)
14
15 -- The laplace mechanism, assuming the server state is
16 -- already given, has type
17 -- (User -> Bool) -> Enclave Double
18 -- In the example usage in the paper, the salaryWithin
19 -- query is partially applied such that it is of type
```

```haskell
20 -- User -> Bool
21 laplaceMechanism :: Enclave (Ref CleanRoomSt) -> (User ->
      Bool) -> Enclave Double
22 laplaceMechanism refst q = do
23    st <- readRef =<< refst
24    -- perform the true query
25    true <- int2Double <$> countingQuery refst q
26    -- sample noise from the laplace distribution
27    noise <- laplaceDistribution refst (1 / (epsilon st))
28    -- augment the true result with the noise
29    return $ true + noise
```

**Listing 12.** Code for the laplace mechanism.